\documentclass[amsmath, amssymb, preprintnumbers, showpacs, showkeys, aps, prb,superscriptaddress,twocolumn]{revtex4-2}
\usepackage{amssymb}
\usepackage[utf8]{inputenc}
\usepackage{graphicx}
\usepackage{amsmath}
\usepackage{csquotes}
\usepackage{xcolor}
\usepackage{placeins}
\usepackage{braket}
\usepackage{comment}
\usepackage[colorlinks, linkcolor={red!80!black},citecolor={blue!70!black},urlcolor={blue!80!black}]{hyperref}

\begin{document}

\title{Spatiotemporal chaos in the interface growth of topological insulators}

\author{Yutaro Tanaka}
\affiliation{
		RIKEN Center for Emergent Matter Science, Wako  351-0198, Japan
	}

\author{Akira Furusaki}
\affiliation{
		RIKEN Center for Emergent Matter Science, Wako  351-0198, Japan
	}

\begin{abstract}
We demonstrate that topological insulators exhibit an intrinsic interfacial instability that amplifies small interface fluctuations, resulting in chaotic behavior during interface growth. 
This mechanism is different from conventional interfacial instabilities in crystal growth that are driven by external non-uniformities such as surface diffusion, and instead arises from intrinsic electronic properties of topological materials.
We find that the boundary states of topological insulators have a pronounced impact on the surface stiffness, which quantifies how strongly a surface resists changes in its shape or orientation. While trivial insulators possess positive stiffness that smooths out surface roughness, topological insulators exhibit negative stiffness that amplifies small shape fluctuations.
We derive an effective equation of the interface growth with this negative stiffness and demonstrate that the interface dynamics is governed by the Kuramoto--Sivashinsky equation, a prototypical nonlinear equation exhibiting spatiotemporal chaos.
\end{abstract}

\maketitle

\section{Introduction}
Understanding interfacial growth between a solid and its surrounding environment, such as vapor or liquid phases, is an important theme in condensed matter physics. 
Interface growth often involves nonlinear dynamics~\cite{Kardar1986} and may exhibit morphological instabilities that lead to chaotic behavior, characterized by irregular dynamics and sensitive dependence on initial conditions in deterministic systems. 
In particular, chaos characterized by irregular dynamics in both space and time is called spatiotemporal chaos~\cite{Cross1993}. Typical examples include the Belousov--Zhabotinsky reaction~\cite{Epstein1996} and the Kuramoto--Sivashinsky equation, which describes chemical reaction systems and flame propagation~\cite{Kuramoto1976, Sivashinsky1977}. 
The interfacial growth of crystals can also exhibit such spatiotemporal chaos in the presence of interfacial instabilities~\cite{Bena1993, Saito1994}, which are typically attributed to extrinsic effects, such as diffusion of adatoms along the solid surface~\cite{Schwoebel1966, Bales1990, Uwaha1992}. 
Furthermore, kinetic mechanisms originating from specific environmental conditions, such as the steering effect caused by long-range attractive forces during grazing incidence deposition, are also known to significantly enhance surface roughening~\cite{Dijken1999PRL, Montalenti2001PRB, Raible2002}.
However, it remains elusive whether intrinsic material properties alone can drive interfacial instabilities.   

Topological insulators host gapless boundary states~\cite{RevModPhys.82.3045, RevModPhys.83.1057}. 
These boundary states can significantly modify the surface energy~\cite{PhysRevLett.129.046802, PhysRevB.107.245148, Mondal2025, tanaka2025fractal}, which is the energy required to create a surface from the bulk crystal and plays a crucial role in determining equilibrium crystal morphology \cite{wulffconst, wulffconst2, PhysRev.82.87, ringe2013kinetic}.
However, despite extensive developments in topological band theory, the role of topological boundary states in nonequilibrium interface dynamics is largely unexplored. 
Experimentally, the growth of the topological insulator thin films, such as $\textrm{Bi}_2\textrm{Se}_3$ and $\textrm{Bi}_2\textrm{Te}_3$, is known to be accompanied by interfacial roughening~\cite{Li2010, Richardella2010, Chen2010, Liu2012, Zeng2013,  Jerng2013}. These morphological features are typically attributed to extrinsic factors, such as substrate morphology and lattice mismatch.

\begin{figure}
\includegraphics[width=1.\columnwidth]{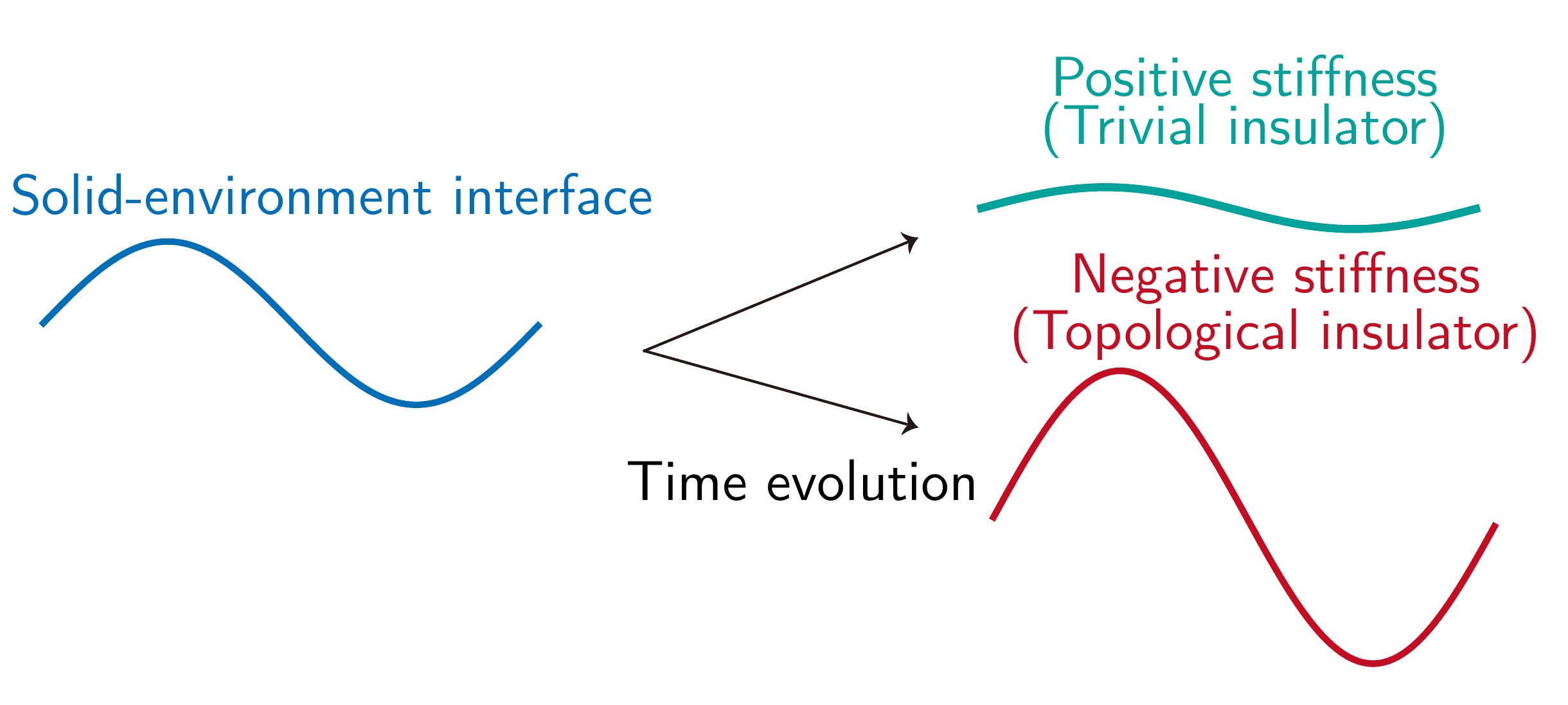}
\caption{Schematic illustration of interface growth with positive and negative stiffness.
}\label{fig:concept}
\end{figure}

In this paper, we demonstrate that topological insulators have an intrinsic interfacial instability that is closely related to their boundary states. 
We show that the boundary states of topological insulators significantly affect the surface stiffness, a measure of a surface's resistance to change in its shape. While positive stiffness smooths out surface roughness, negative stiffness destabilizes a flat surface and amplifies small shape fluctuations instead of smoothing them out~[Fig.~\ref{fig:concept}]. 
We demonstrate that the boundary states of topological insulators induce negative stiffness.
Furthermore, we show that the interface growth with this negative stiffness in topological insulators is described by the Kuramoto--Sivashinsky equation, a prototypical nonlinear partial differential equation known to exhibit spatiotemporal chaos.  
This mechanism is intrinsic to the topological electronic structure and, in stark contrast to conventional instability mechanisms in crystal growth.

\section{Surface energy of band insulators}
For simplicity, we consider the ground state of noninteracting electrons at half filling on a two-dimensional (2D) centrosymmetric lattice, where the lattice constant is set equal to be 1.

The ground-state energy per unit cell of a bulk 2D insulator is given by
\begin{align}\label{eq:bulk_enrgy}
    E_{\rm bulk}=\frac{1}{(2\pi)^2}\sum_{n}\int_{\rm BZ}E^{(n)}_{{\rm PBC},\boldsymbol{k}} \Theta \!\left( -E^{(n)}_{{\rm PBC, \boldsymbol{k}}}\right)d^2 k,
\end{align}
where $E^{(n)}_{{\rm PBC},\boldsymbol{k}}$ is the energy eigenvalue for the $n$th band under the full periodic boundary conditions (PBCs), and $\Theta(x)$ is the Heaviside step function. 
We have set the Fermi energy to be zero.

\begin{figure}
\includegraphics[width=1.\columnwidth]{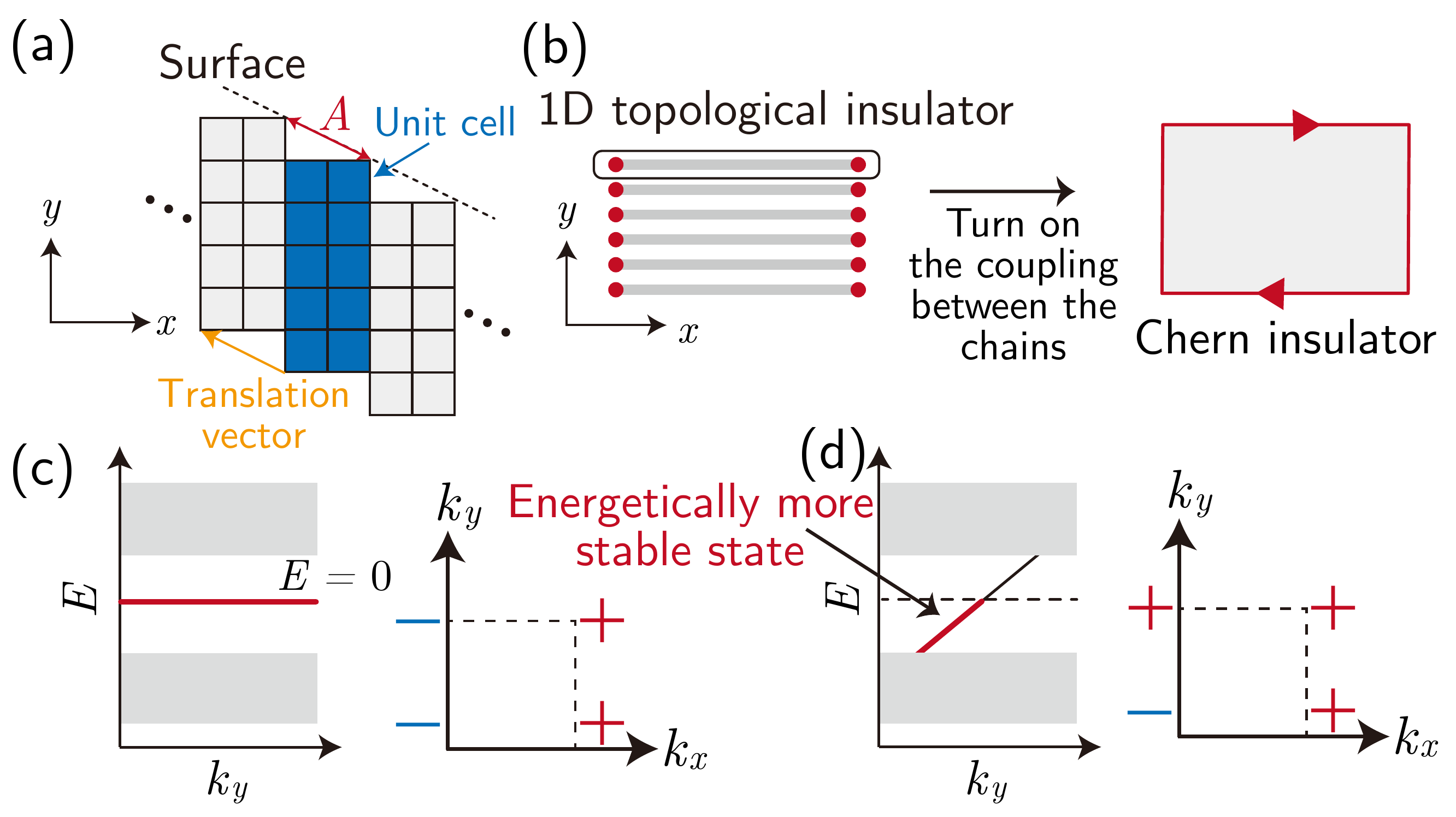}
\caption{(a)~Schematic illustration of a square lattice in strip geometry. The blue shaded region represents the unit cell in this geometry, and $A$ is the surface length of the unit cell. (b)~The transition from decoupled 1D topological insulators to a Chern insulator. (c) and (d)~The band structures of the boundary states along the $y$ direction (left panel) and parity eigenvalues at the TRIM (right panel) (c) for the decoupled 1D topological insulators and (d) for the Chern insulator. The red lines in the band structures indicate the occupied boundary states. 
}\label{fig:concept2}
\end{figure}

Next we consider the energy of the Hamiltonian $H(k_{\parallel})$ under the open boundary condition (OBC) in one direction and the PBC in the other direction, where $k_{\parallel}$ is the wavevector along the surface. 
We refer to this geometry as the strip geometry. 
The energy in the strip geometry is given by  
\begin{align}\label{eq:energy_slab}
    E_{\rm strip}=\frac{1}{2\pi}\sum_{n}\int^{\pi}_{-\pi}E^{(n)}_{{\rm OBC}, {k}_{\parallel}}\Theta \!\left( -E^{(n)}_{{\rm OBC}, {k}_{\parallel}} \right)d{k}_{\parallel},
\end{align}
where $E^{(n)}_{{\rm OBC}, {k}_{\parallel}}$ is the $n$th energy eigenvalue of the Hamiltonian $H(k_{\parallel})$. 
By using these energies, we define the surface energy for systems with spatial-inversion symmetry by~\cite{Boettger1994, Vincenzo1996, tran2016surface}
\begin{align}\label{eq:surf_ene}
    \gamma := \frac{E_{\rm strip}-N_{\rm uc}E_{\rm bulk}}{2A},
\end{align}
where $N_{\rm uc}$ is the number of unit cells of the bulk system contained in the larger unit cell of $H(k_\parallel)$ in the strip geometry, and $A$ is the surface length of the unit cell in the strip geometry. 
For example, $N_{\rm uc}=10$ and $A=\sqrt{3}$ in the strip geometry shown in Fig.~\ref{fig:concept2}(a).
The number of occupied subbands in Eq.\ (\ref{eq:energy_slab}) equals $N_{\rm uc}$ times the number of occupied bands in Eq.\ (\ref{eq:bulk_enrgy}).
In the absence of inversion symmetry, the two surfaces of a cleaved crystal are no longer equivalent, which leads to an intrinsic ambiguity in defining the surface energy~\cite{ARBEL1975305,LEE1975302,wang2022defining}.
To avoid this complication, we confine our analysis to systems with inversion symmetry throughout this work.

We show that the boundary states of topological insulators play a crucial role in determining the surface energy.
Here we take a Chern insulator as a representative example of topological insulators without time-reversal symmetry, but our theory is also applicable to topological insulators with time-reversal symmetry, as discussed in Appendix~\ref{appendix: TRS}. 
We construct a Chern insulator by stacking one-dimensional (1D) centrosymmetric topological insulators that respect chiral symmetry $\Gamma H \Gamma^{-1}=-H$ with $\Gamma$ being a unitary matrix satisfying $\Gamma^2=1$~[Fig.~\ref{fig:concept2}(b)]. 
Each 1D topological insulator is characterized by the Zak phase~\cite{Zak1989}, defined as a line integral over the 1D Brillouin zone, 
\begin{align}
    \vartheta = -i \sum_{n}^{\rm occ.} \int^{2\pi}_{0}dk_x \bra{u_n} \frac{\partial}{\partial {k_x}}\ket{u_n}, 
\end{align}
where $u_n$ is the periodic part of the Bloch wave function of the $n$th band, and the summation is taken over the occupied bands. In the presence of inversion symmetry, the Zak phase is quantized and takes the values $0$ or $\pi$ (mod $2\pi$). 

We consider a 2D system composed of the decoupled 1D topological insulators that host zero-energy modes localized at their edges~[Fig.~\ref{fig:concept2}(b)] owing to the nontrivial Zak phase $\vartheta=\pi$~\cite{Ryu2002}. 
Since we focus on systems with inversion symmetry, we can evaluate the Zak phase from the parity eigenvalues at the time-reversal invariant momenta (TRIM)~\cite{Hughes2011PRB} via 
\begin{align}
     e^{i\vartheta(k_y)}= \prod_{n}^{\rm occ.}\xi_{n}(k_x=0,k_y)\xi_{n}(k_x=\pi,k_y), 
\end{align}
where $k_y=0$ or $\pi$, and $\xi_{n}(k_x)$ denotes the parity eigenvalue of the $n$th band.
Figure~\ref{fig:concept2}(c) shows a representative configuration of parity eigenvalues at the TRIM that gives rise to the nonzero Zak phase $\vartheta=\pi$ for both $k_y=0$ and $k_y=\pi$. 
Furthermore, the parity of the Chern number $C$ can be determined from the parity eigenvalues~\cite{Hughes2011PRB, PhysRevB.86.115112} via
\begin{align}
     (-1)^{C}= \prod_{n}^{\rm occ.}\xi_{n}(\Gamma)\xi_{n}(X)\xi_{n}(Y)\xi_{n}(M), 
\end{align}
where $\xi_{n}(\Gamma_i)$ is the parity eigenvalue of the $n$th occupied band at the TRIM, $ \Gamma=(0,0)$, $X=(\pi,0)$, $Y=(0,\pi)$, and $M=(\pi,\pi)$.
Figure~\ref{fig:concept2}(d) also illustrates a representative configuration of parity eigenvalues that yields a nonzero Chern number.

When the band inversion occurs at $\boldsymbol{k}=Y$, the system transforms from a set of 1D topological insulators into a Chern insulator. 
To induce this band inversion, we introduce the coupling between the 1D topological insulators. 
Specifically, we define a real parameter $g$ that controls the strength of the coupling in such a way that the system is a set of decoupled 1D insulators at $g=0$ and becomes a Chern insulator at $g=1$.
A specific lattice model is introduced in Eq.\ (\ref{eq:Chern_model}) below.

Let us consider the variation in the surface energy $\gamma$ of the (10) surface as $g$ increases from $g=0$ to $g=1$.
In trivial insulators, we expect that the surface energy $\gamma$ should be primarily determined by the energy associated with the dangling bonds on the (10) surface~\cite{jeong1999steps, galanakis2002SS, galanakis2002EPL, PhysRevLett.92.086102}.
In contrast, in topological insulators, the boundary states are expected to make an additional contribution to the surface energy. 
As $g$ is varied from $0$ to $1$, the zero-energy boundary states with the flat dispersion $E=0$ turn into a chiral boundary mode with linear dispersion, $E=v k_y$, where the velocity $v$ is a real constant. 
Thus, the energy change of the boundary states, $\Delta E$, associated with the change in $g$ from $0$ to $1$, 
is given by
\begin{align}\label{eq:boundary_energy}
    \Delta E= v \int_{-k_c}^{0} k_y dk_y <0,
\end{align}
where $k_c$ is a cutoff wavevector.
As we will demonstrate below, this energy change $\Delta E$ significantly contributes to the surface energy $\gamma$ and modifies the dynamics of the surface growth of Chern (topological) insulators.

Incidentally, we note that, in general, the surface energy is not determined solely by the contribution of the boundary states. As discussed in Sec.~\ref{sec: dep_surf_ene}, the dangling bonds at the surface significantly contribute to this energy, resulting in a net positive surface energy. In contrast, the surface stiffness, which is a distinct quantity defined in Eq.~(\ref{eq:stiffness}), becomes negative due to the presence of the boundary states, as we will discuss in Sec.~\ref{sec: interface_growth}.

\begin{figure}
\includegraphics[width=1.\columnwidth]{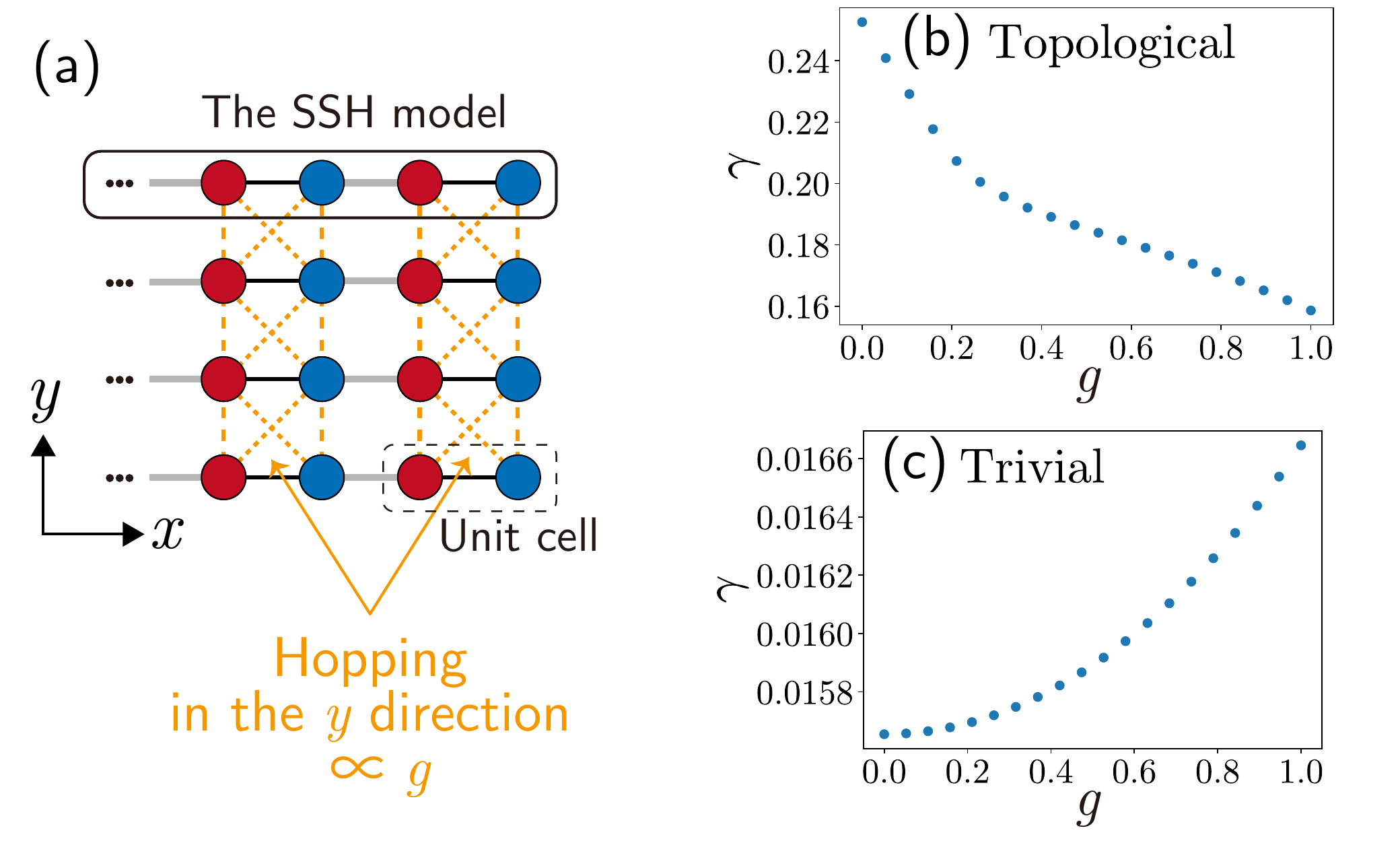}
\caption{
(a)~Schematic illustration of the tight-binding model in Eq.~\eqref{eq:Chern_model}. The hopping amplitude in the $y$ direction is proportional to the real parameter $g$. In the absence of hopping in the $y$ direction, this model reduces to a set of decoupled SSH chains. (b) and (c)~The surface energy $\gamma$ for the (10) surface for (b) $m=0.8$ and (c) $m=4$ in the model~[Eq.~\eqref{eq:Chern_model}]. 
The parameter is set to $v=0.7$, and the system size is $L_x=L_y=40$.
}\label{fig:energy_vs_lambda}
\end{figure}

To investigate the contribution of the boundary states to the surface energy, we introduce a simple tight-binding model of a Chern insulator~\cite{PhysRevB.74.085308} on a square lattice with the lattice sites denoted by a lattice vector $\boldsymbol{R}=(x,y)$ with $x \in \mathbb{Z}$ and $y \in \mathbb{Z}$:
\begin{align}
    H_{g}(\boldsymbol{k})={}&(m-\cos k_x - g \cos k_y )\sigma_x \nonumber \\
    &+ v(\sin k_x \sigma_y +g \sin k_y \sigma_z), \label{eq:Chern_model}
\end{align}
where $\sigma_i$ ($i=x, y, z$) are the Pauli matrices, and $m, v, g \in \mathbb{R}$.
In this model, the unit cell has two sites~[the red and blue sites in Fig.~\ref{fig:energy_vs_lambda}(a)], labeled by $\sigma_z=\pm1$.
The amplitude of the hopping in the $y$ direction is proportional to the parameter $g$ ($0\leq g \leq 1$)~[Fig.~\ref{fig:energy_vs_lambda}(a)]. 
When $g=0$, i.e., when the hopping in the $y$ direction vanishes, the model is reduced to the form
\begin{align}
    H_{0}(k_x)=\begin{pmatrix}
        0 & q(k_x) \\
        q^{*}(k_x) & 0
    \end{pmatrix},
\end{align}
with $q(k_x)=m-\cos k_x -iv \sin k_x$.
This is the Su-Schrieffer-Heeger (SSH) model~\cite{PhysRevLett.42.1698}, a prototypical model for a 1D topological insulator. 
When $1-g<|m|<1+g$, this model hosts a nonzero Chern number and supports a chiral edge mode (see Appendix~\ref{appendix: phase_diagram} for the topological phase diagram of the model).
We solve this model using the PythTB package~\cite{Coh_Vanderbilt_PythTB_2022} and calculate the surface energy $\gamma$.

The surface energy $\gamma$ of our model $H_{g}(\boldsymbol{k})$ for the (10) surface decreases as $g$ increases in the topologically nontrivial phase~($m=0.8$) [Fig.~\ref{fig:energy_vs_lambda}(b)], whereas $\gamma$ increases with $g$ in the trivial phase~($m=4$) [Fig.~\ref{fig:energy_vs_lambda}(c)]. 
Thus, the surface energy $\gamma$ of the Chern insulator ($g=1$) is lower than its value at $g=0$, in clear contrast to the behavior in the trivial insulator phase (see Appendix~\ref{appendix: surface_energy_TPT} for more details on the dependence of the surface energy on $g$ and $m$). 
This result is consistent with our discussion that the boundary states give a negative contribution [Eq.\ (\ref{eq:boundary_energy})] to the surface energy.
From now on, we will fix the parameter $g$ to 1, ensuring that our model $H_{g}(\boldsymbol{k})$ respects four-fold rotation symmetry with respect to the $z$ axis: $C_{4}H_{g}(\boldsymbol{k})C_{4}^{-1}=H_g(k_y,-k_x)$ with $C_{4}=(1-i\sigma_x)/\sqrt{2}$. 

\section{Dependence of the surface energy on the orientation}\label{sec: dep_surf_ene}
\begin{figure}
\includegraphics[width=1.\columnwidth]{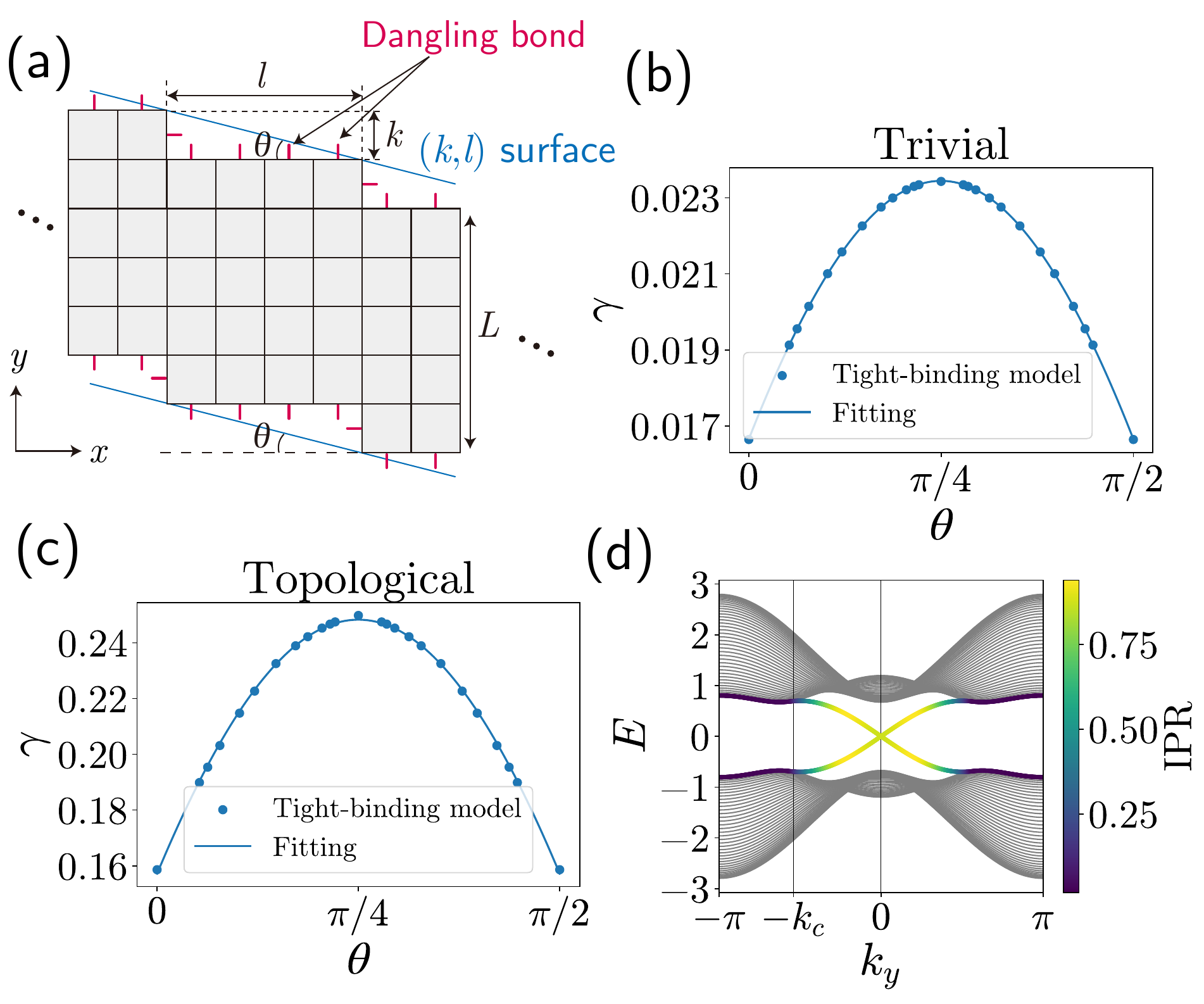}
\caption{(a)~Dangling bonds on the surface with the Miller index $(k,l)$. Each plaquette is a unit cell of a bulk. (b,c)~Surface energy $\gamma$~[Eq.~\eqref{eq:surf_ene}] of the model [Eq.~\eqref{eq:Chern_model}] (b) in the trivial phase ($m=4$) and (c) in the topological phase ($m=0.8$).  ``Tight-binding model" indicates the direct calculation of the model. ``Fitting" indicates the fitting with Eq.~\eqref{eq:broken_bond_ec}. 
(d)~Band structures of the tight-binding model ($m=0.8$) under the OBC in the $x$ direction and the PBC in the $y$ direction. The color bar indicates the inverse participation ratio (IPR) $\sum_{x}|\psi_{x}(k_y)|^4/(\sum_{x}|\psi_{x}(k_y)|^2)^2$ for the boundary states, which characterizes the localization of the boundary states along the two edges in the strip geometry. The parameters are the same as Fig.~\ref{fig:energy_vs_lambda}, and the width of the strip [$L$ in (a)] is 40. 
}\label{fig:surf_ene}
\end{figure}

We discuss the dependence of the surface energy $\gamma$ on the surface orientation based on the broken-bond rule~\cite{jeong1999steps, galanakis2002SS, galanakis2002EPL, PhysRevLett.92.086102}.
According to this rule, the surface energy is dominated by a contribution that is proportional to the number of dangling bonds on the surface. 
Let us consider the $(kl)$ surface inclined at an angle $\theta$ ($0\leq \theta \leq \pi/2$) to the $x$ axis~[Fig.~\ref{fig:surf_ene}(a)]. 
For the $(kl)$ surface, there are $l+k$ dangling bonds in a surface segment of length $A=\sqrt{l^2+k^2}$. 
Hence, the dangling-bond contribution to the surface energy is given by $\gamma_0(\theta)=  e_b(l + k)/\sqrt{l^2+k^2}$, where $e_b$ is the energy associated with a dangling bond oriented along the $x$ or $y$ direction per unit cell; see Fig.~\ref{fig:surf_ene}(a).
Here, the energy per dangling bond in the $x$ direction is equal to that in the $y$ direction because our model respects four-fold rotation symmetry. 
When $0\leq \theta \leq \pi/2$, the relations $\cos \theta = l/\sqrt{l^2+k^2}$ and $\sin \theta = k/\sqrt{l^2+k^2}$ hold.
Thus, the surface energy can be expressed as
\begin{align}\label{eq:broken_bond_ec}
\gamma(\theta) = e_{b} (|\cos \theta| + |\sin \theta|) + \alpha,
\end{align}
where the angular domain is extended to $-\pi \leq \theta \leq \pi$ by replacing $\cos \theta$ and $\sin \theta$
with $|\cos \theta|$ and $|\sin \theta|$, respectively.
Furthermore, we have introduced the constant term $\alpha$, which will be shown below to play a crucial role in surface growth.

Figures~\ref{fig:surf_ene}(b) and \ref{fig:surf_ene}(c) show the dependence of the surface energy $\gamma$ of the model~[Eq.~\eqref{eq:Chern_model}] on the angle $\theta$ in the trivial and topological phases, respectively. 
Here, we focus on the region $0\leq \theta \leq \pi/2$ since our model respects the four-fold rotation symmetry. 
We fit the surface energy of both the trivial and topological phases using Eq.~\eqref{eq:broken_bond_ec}
and obtain $e_b=0.0164$ and $\alpha=0.000265$ for (b) and $e_b=0.220$ and $\alpha = -0.0627$ for (c). 
In the trivial phase, $\alpha$ takes a small positive value, indicating that the surface energy $\gamma$ is almost entirely governed by the broken-bond rule.
This is because the present calculation considers the ground state of a simple tight-binding model.
If finite-temperature effects, such as step–step interactions~\cite{GRUBER1967875, Jayaprakash1984}, are introduced, $\gamma$ would acquire additional contributions beyond the broken-bond rule. 

Interestingly, in the topological phase, $\alpha$ is negative, and its magnitude is not very small compared with $e_b$.
The negative $\alpha$ originates from the presence of boundary states.
As discussed below Eq.\ (\ref{eq:boundary_energy}), the boundary states give a negative contribution to the surface energy, in addition to the energy cost of broken bonds when forming the surface from the bulk.
This contribution from the boundary states is constant regardless of the angle $\theta$.
Thus, we expect $\alpha$ to be related to the energy eigenvalues of the boundary states.
The above value of $\alpha=-0.0627$ can indeed be accounted for by
\begin{align}
        \alpha \simeq \frac{1}{2\pi}\int^{0}_{-k_c}E_{\textrm{edge}}(k_{y})\Theta \left( -E_{\textrm{edge}}(k_{y}) \right)dk_{y}, \label{eq: alpha_energy}
\end{align}
where $E_{\rm edge}(k_y)$ is the energy of the boundary mode colored in Fig.~\ref{fig:surf_ene}(d), and the cutoff wavevector $k_c$ is chosen as $k_c=1.7$.

Finally, we note that real crystal surfaces may deviate from the broken-bond rule due to surface reconstructions. While such reconstructions can quantitatively modify the dangling-bond contribution to $\gamma$, the negative contribution $\alpha$ from the topological boundary states [Eq.~\eqref{eq: alpha_energy}] arises independently of the specific form of the dangling-bond contribution. Consequently, the interfacial instability driven by $\alpha$, as discussed in Sec.~\ref{sec: interface_growth}, is expected to persist as long as the boundary states remain gapless.

\section{Interface growth with the negative stiffness}\label{sec: interface_growth}
We introduce the surface stiffness $ \tilde{\gamma}(\theta)$, which quantifies how strongly a surface resists changes in its shape. From the surface energy in Eq.~\eqref{eq:broken_bond_ec}, we can see that the surface stiffness~\cite{Pimpinelli_Villain_1998} is equal to the constant $\alpha$,
\begin{align}\label{eq:stiffness}
    \tilde{\gamma}(\theta) := \gamma(\theta) + \frac{\partial^2 \gamma}{\partial \theta^2}=\alpha, \quad (|\theta|\neq 0, \pi/2, \pi),
\end{align}
where we exclude the high-symmetry orientations $|\theta|=0, \pi/2$, and  $\pi$ from the stiffness calculation, since the broken-bond rule exhibits non-analytic cusps at these angles where the surface stiffness formally diverges.
The positive stiffness ($\tilde{\gamma}=\alpha > 0$) for the trivial phase smooths out surface roughness, whereas the negative stiffness ($\tilde{\gamma}=\alpha<0$) for the topological phase amplifies small shape fluctuations.
We confirm that the Chern insulator in Eq.~\eqref{eq:Chern_model} has a lower surface energy for a bumped surface than for a flat one, which is consistent with negative stiffness, as discussed in Appendix~\ref{appendix: interfacial_instability}. 

\begin{figure}
\includegraphics[width=1.\columnwidth]{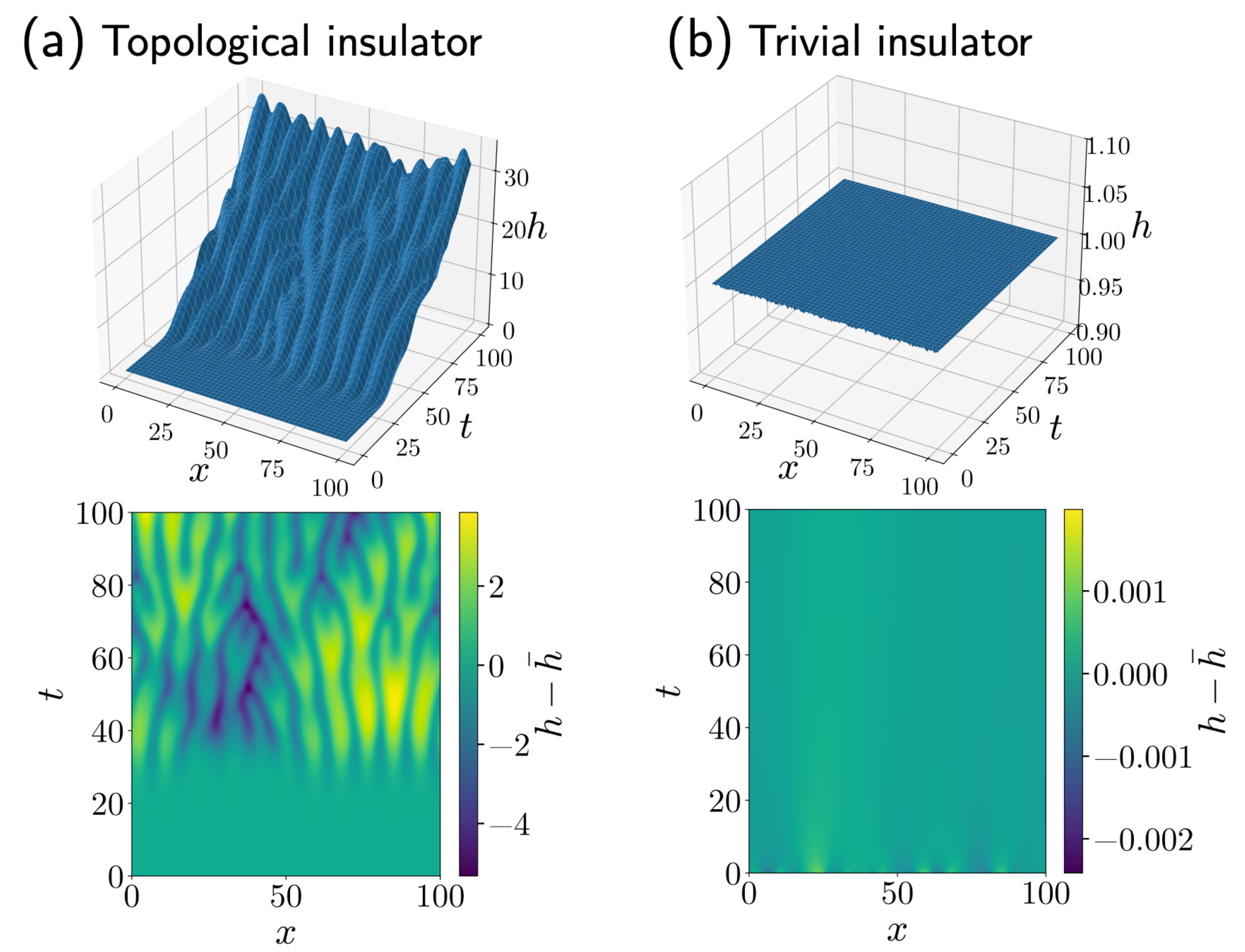}
\caption{The dynamics of the interface $h$ described by Eq.~\eqref{eq:interface_growth} with the system size $L=100$ under the PBC for (a) the topological insulator~($\tilde{\gamma}<0$) and (b) the trivial insulator~($\tilde{\gamma}>0$). 
The lower panels show the spatiotemporal evolution of the interface fluctuation $h-\bar{h}$. 
Here, $\bar{h}$ denotes the spatial average of the interface height $h$ at each time step [$\nu=-1$ in (a) and $\nu=1$ in (b). $0\leq t\leq 100$, $\kappa=1$, and $\eta=1$ in both (a) and (b)].
}\label{fig:interface_growth}
\end{figure}

We analyze how negative stiffness influences the dynamics of an interface $ h(x,t)$, which depends on position $ x $ and time $ t $, for a slightly tilted surface ($0<\theta \ll 1 $).
Here, we assume the kinetic-limited growth of a solid from a surrounding uniform vapor or liquid phase, where the growth is driven by the chemical potential difference between the environment and the solid~\cite{saito1996book, Pimpinelli_Villain_1998}.
As derived in Appendix~\ref{appendix: derivation_interfaace_growth}, when the interface growth is governed by a thermodynamic driving force, the time evolution of the interface is given by
\begin{align}\label{eq:interface_growth}
    \frac{\partial h}{\partial t}
    = \nu\frac{\partial^2 h}{\partial x^2}
    + \kappa \left( \frac{\partial h}{\partial x} \right)^2
    - \eta \frac{\partial^4 h}{\partial x^4},
\end{align}
where $\nu$ is a coefficient related to the surface tension and positive for $\tilde{\gamma}>0$ and negative for $\tilde{\gamma}<0$, $\kappa$ ($> 0$) is a coefficient related to the interface growth velocity, and $\eta$ ($> 0$) is one related to the effect suppressing the interfacial roughening (see Appendix~\ref{appendix: derivation_interfaace_growth} for detailed definitions of these coefficients).
The nonlinear term in Eq.~\eqref{eq:interface_growth} arises naturally because the interface advances locally along its normal direction, rather than in the vertical ($h$) direction. Projecting this normal growth velocity onto the vertical axis geometrically introduces the nonlinear term~\cite{Kardar1986}.
The fourth-order derivative term in Eq.~\eqref{eq:interface_growth} arises from the elastic free energy of an undulating solid surface. The energy scale of this elastic contribution is typically one to two orders of magnitude smaller than that of the surface energy.
Therefore, in conventional interface growth, it acts merely as a higher-order correction. In contrast, when $\nu$ is negative, the leading second-order derivative term drives an interfacial instability. In the absence of this fourth-order term, the equation yields infinitely unstable and unphysical solutions. Thus,  the inclusion of the fourth-order derivative term is essential to obtain stable and physical solutions.

Equation~\eqref{eq:interface_growth} with a negative $\nu$ ($\tilde{\gamma}<0$) is the Kuramoto--Sivashinsky equation~\cite{Kuramoto1976, Sivashinsky1977} exhibiting spatiotemporal chaos. 
In the topological phase, the negative stiffness $ \tilde{\gamma}$ directly yields this negative coefficient.
To elucidate the dynamics described by this equation, we substitute the Fourier expansion $h(x,t)=\sum_{k}h_{k}(t)e^{ikx}$ into Eq.~\eqref{eq:interface_growth} to obtain
\begin{align}\label{eq:KS_fourier}
    \frac{d h_k}{dt}=-\!\left( \nu k^2 + \eta k^4 \right)h_k+\kappa \sum_{k'}k'(k'-k)h_{k-k'}h_{k'}.
\end{align}
We first consider this equation in the absence of the nonlinear term ($\kappa=0$) and then obtain
\begin{align}\label{eq:liner_solution}
    h_{k}(t)= \exp\!\left[ {-\left( \nu k^2 + \eta k^4 \right) t} \right] h_{k}(0).
\end{align}
When $\nu<0$, the long-wavelength modes ($|k|<\sqrt{-\nu/\eta}$) are linearly unstable, whereas the short-wavelength modes ($|k|>\sqrt{-\nu/\eta}$) are stabilized.
In the presence of the nonlinear term, nonlinear coupling induces competition among the Fourier modes, which eventually gives rise to spatiotemporal chaos~\cite{Papageorgiou1991}. 

Figure~\ref{fig:interface_growth}(a) shows the results of numerical simulations of the time evolution governed by Eq.~\eqref{eq:interface_growth} for a negative value of $\nu$. 
The initial condition is a nearly flat interface with small random perturbations, $h(x,0)=1+10^{-3}\epsilon(x)$, where $\epsilon(x)$ is Gaussian noise with zero mean and unit variance.
As the growth proceeds, the interface fluctuation $h-\bar{h}$ develops into a chaotic state characterized by irregular spatiotemporal variations, where $\bar{h}$ is the spatial average of $h$.
In contrast, for a positive value of $\nu$, the initial perturbations are smoothed out, and no chaotic behavior emerges~[Fig.~\ref{fig:interface_growth}(b)], because the second-order derivative term in Eq.~\eqref{eq:interface_growth} acts as a positive diffusion that stabilizes long-wavelength modes.
This contrast highlights that the sign of the effective surface stiffness $\tilde{\gamma}$ controls the stability of the interface morphology.

\section{Discussion}

In summary, we have elucidated the effects of the boundary states on the interfacial growth of the 2D topological insulators with and without time-reversal symmetry. We have shown that topological insulators have an intrinsic interfacial instability characterized by negative surface stiffness, which amplifies the small shape fluctuations. Furthermore, we derived an equation that describes the interface dynamics of the topological insulators and demonstrated that it reduces to the Kuramoto--Sivashinsky equation describing spatiotemporal chaos. 

Although we have focused on 2D systems in this paper, our theory can be naturally extended to three-dimensional systems. We have shown that the boundary states lead to negative stiffness by considering the transition from 1D topological insulators to a 2D topological insulator. Since the same argument applies to the transition from stacked 2D topological insulators to a three-dimensional topological insulator, we expect that three-dimensional topological insulators also exhibit negative stiffness. 
Our theory offers a novel perspective on the crystal morphology of the topological insulators.

\begin{acknowledgments}
Y.T. thanks Shuichi Murakami and Kohei Kawabata for fruitful discussions. This work was supported by Japan Society for the Promotion of Science (JSPS) KAKENHI Grant No.~JP24K22868 and by JST CREST Grant No.~JPMJCR19T2. 
\end{acknowledgments}

\appendix

\section{Topological insulators with time-reversal symmetry}\label{appendix: TRS}
Here, we discuss the surface stiffness of a topological insulator with time-reversal symmetry.
A 2D topological insulator with time-reversal symmetry can be reduced to a set of 1D topological insulators by turning off the hopping amplitudes along one direction, in a manner analogous to the case of the Chern insulator. Therefore, a 2D topological insulator with time-reversal symmetry is also expected to exhibit negative stiffness, similarly to a Chern insulator. 
In the following, we demonstrate this using the Bernevig--Hughes--Zhang~(BHZ) model~\cite{bernevig2006quantum}, which is a prototypical model of a 2D topological insulator with time-reversal symmetry. 
This model is defined on a square lattice with the lattice sites denoted by a lattice vector $\boldsymbol{R}=(x,y)$ with $x\in \mathbb{Z}$ and $y\in \mathbb{Z}$. The Bloch Hamiltonian is given by
\begin{align}\label{eq:BHZmodel}
    H(\boldsymbol{k})=&\left(m - \cos k_x - \cos k_y \right) \tau_z \nonumber \\
    &+v(\sin k_x \tau_y + \sin k_y \sigma_z \tau_x),
\end{align}
where $\tau_i$ ($i=x,y,z$) is the Pauli matrices and $m, v \in \mathbb{R}$.
This model respects time-reversal symmetry $\mathcal{T}H(\boldsymbol{k})\mathcal{T}^{-1}=H(-\boldsymbol{k})$ with $\mathcal{T}=i\sigma_y K$, where $K$ denotes the complex conjugation operator. 

\begin{figure}
\includegraphics[width=1.\columnwidth]{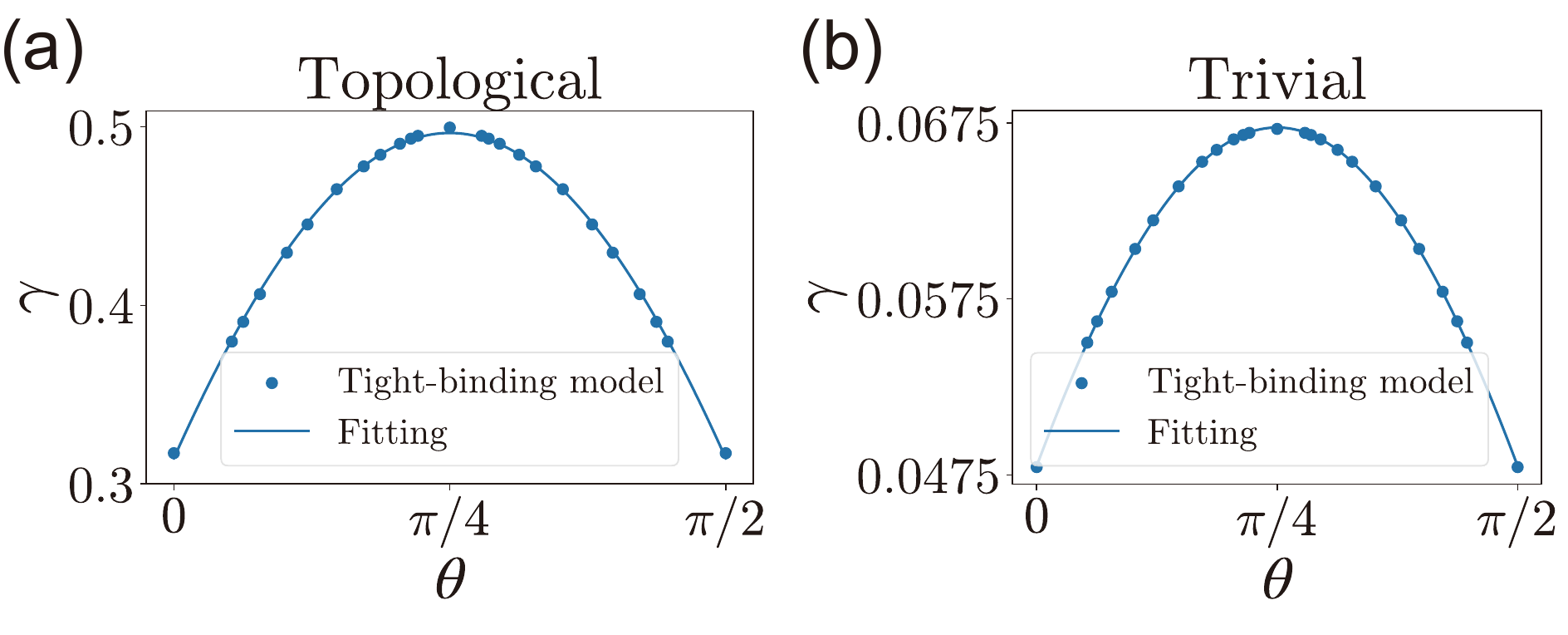}
\caption{Surface energy $\gamma$~[Eq.~\eqref{eq:surf_ene}] of the BHZ model [Eq.~\eqref{eq:BHZmodel}] with $v=0.7$ (a) in the topological phase ($m=0.8$) and (b) in the trivial phase ($m=3$). ``Tight-binding model" indicates the direct calculation of the model. ``Fitting" indicates the fitting with Eq.~\eqref{eq:broken_bond_ec}. 
The width of the strip is 40. 
}\label{fig:BHZ}
\end{figure}

Figures~\ref{fig:BHZ}(a) and \ref{fig:BHZ}(b) show the surface energy defined in Eq.~\eqref{eq:surf_ene} for the BHZ model in the topological phase~($m=0.8$) and trivial phase~($m=3$), respectively, where $\theta$ denotes the angle between the surface and the $x$ axis.
We fit the surface energy in Fig.~\ref{fig:BHZ} using Eq.~\eqref{eq:broken_bond_ec} in the same manner as for the Chern insulator. 
The obtained values of the fitting parameters are  $e_b=0.440$ and $\alpha=-0.125$ for (a), and $e_b=0.0464$ and $\alpha = 0.00157$ for (b). 
Since the surface stiffness $\tilde{\gamma}$ is equal to $\alpha$ from Eq.~\eqref{eq:stiffness} for $|\theta|\neq0,\pi/2, \pi$, it follows that $\tilde{\gamma}$ is negative in the topological phase and positive in the trivial phase. 

\section{Phase diagram}\label{appendix: phase_diagram}
In this section, we present the topological phase diagram of the tight-binding model introduced in the main text. 
The Hamiltonian is given by
\begin{align}\label{eq:SM_model}
    H_{g}(\boldsymbol{k}) = \boldsymbol{R}(\boldsymbol{k}) \cdot \boldsymbol{\sigma},
\end{align}
where $m,v,g \in \mathbb{R}$ are real parameters with $0\leq g \leq 1$, and $\boldsymbol{\sigma}=(\sigma_x, \sigma_y, \sigma_z)^{T}$ is the vector of the Pauli matrices. 
The vector $\boldsymbol{R}(\boldsymbol{k})$ is defined as
\begin{align}
    \boldsymbol{R}(\boldsymbol{k})= \begin{pmatrix}
        R_x(\boldsymbol{k})\\
        R_y(\boldsymbol{k}) \\
        R_z(\boldsymbol{k})
    \end{pmatrix}
    = \begin{pmatrix}
        m - \cos k_x -g \cos k_y\\
        v \sin k_x \\
        v g \sin k_y
    \end{pmatrix}.
\end{align}
The energy eigenvalues are given by
\begin{align}
    E=\pm\sqrt{\sum_{i=x,y,z}R_i^2(\boldsymbol{k})}.
\end{align}

\begin{figure}
	\includegraphics[width=1.\columnwidth]{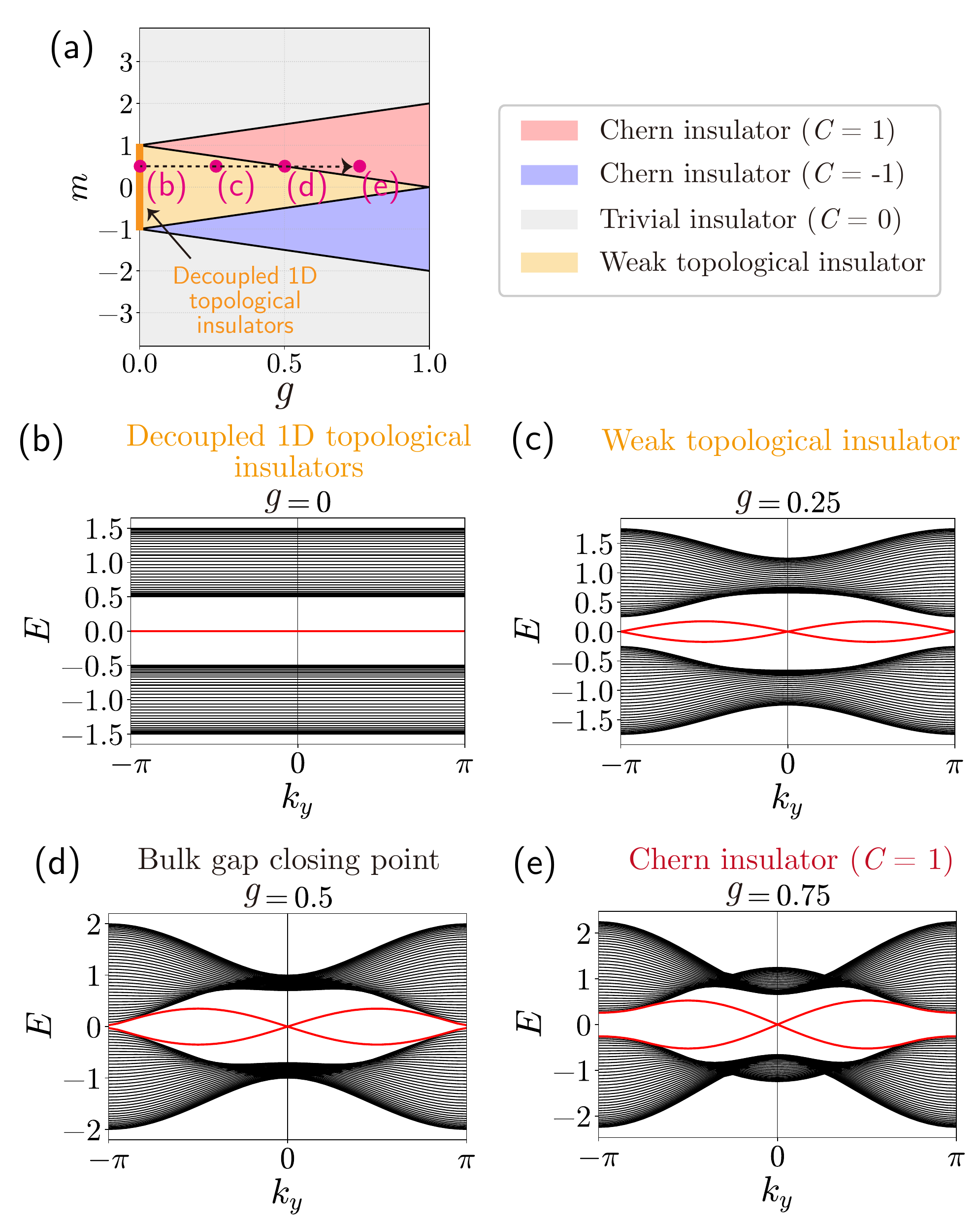}
	\caption{(a)~Topological phase diagram of the model in Eq.~\eqref{eq:SM_model}. (b)--(e)~Band structures of the model under the periodic-boundary condition in the $y$ direction and the open-boundary condition in the $x$ direction for (b) $g=0$, (c) $g=0.25$, (d) $g=0.5$, and (e) $g=0.75$. The parameters are set to $m=0.5$, $v=0.7$, and the system size is $L_x=L_y=40$.}
	\label{fig:phase_diagram}
\end{figure} 

The band gap closes ($E=0$) when the parameters $g$ and $m$ satisfy $R_x(\boldsymbol{k})=R_y(\boldsymbol{k})=R_z(\boldsymbol{k})=0$. 
The components $R_y(\boldsymbol{k})$ and $R_z(\boldsymbol{k})$ vanish only at TRIM $\Gamma=(0,0)$, $X=(\pi,0)$, $Y=(0,\pi)$, and $M=(\pi,\pi)$, 
and therefore the band gap closes only at the TRIM.
The condition $R_x(\boldsymbol{k})=0$ at the TRIM leads to 
\begin{gather}
    m=g+1, \label{eq:SM_gapclose1}\\
    m=g-1, \label{eq:SM_gapclose2} \\
    m=-g+1, \label{eq:SM_gapclose3}\\
    m=-g-1, \label{eq:SM_gapclose4} 
\end{gather}
at the $\Gamma$, $X$, $Y$, and $M$ points, respectively.
Consequently, the band gap closes along the phase boundaries defined by Eqs.~\eqref{eq:SM_gapclose1},  \eqref{eq:SM_gapclose2}, \eqref{eq:SM_gapclose3} or  \eqref{eq:SM_gapclose4}.
To characterize the topological phase, we calculate the Chern number 
\begin{gather}
    C=\frac{1}{2\pi}\int_{\rm BZ}d^2 k \left[ \nabla_{\boldsymbol{k}}\times \boldsymbol{a}(\boldsymbol{k}) \right]_z,
\end{gather}
in the regions bounded by these phase boundaries, where $\boldsymbol{a}(\boldsymbol{k}):=-i\bra{u_{\boldsymbol{k}}}\nabla_{\boldsymbol{k}}\ket{u_{\boldsymbol{k}}}$ is the Berry connection of the occupied band, and $\ket{u_{\boldsymbol{k}}}$ is the periodic part of the Bloch wavefunction.
Based on these results, we obtain the topological phase diagram of the model shown in Fig.~\ref{fig:phase_diagram}(a).
As shown in Fig.~\ref{fig:phase_diagram}(a), we find that for $g-1<|m|<g+1$, the Chern number is given by $C=\pm1$. 

\begin{figure}
	\includegraphics[width=1.\columnwidth]{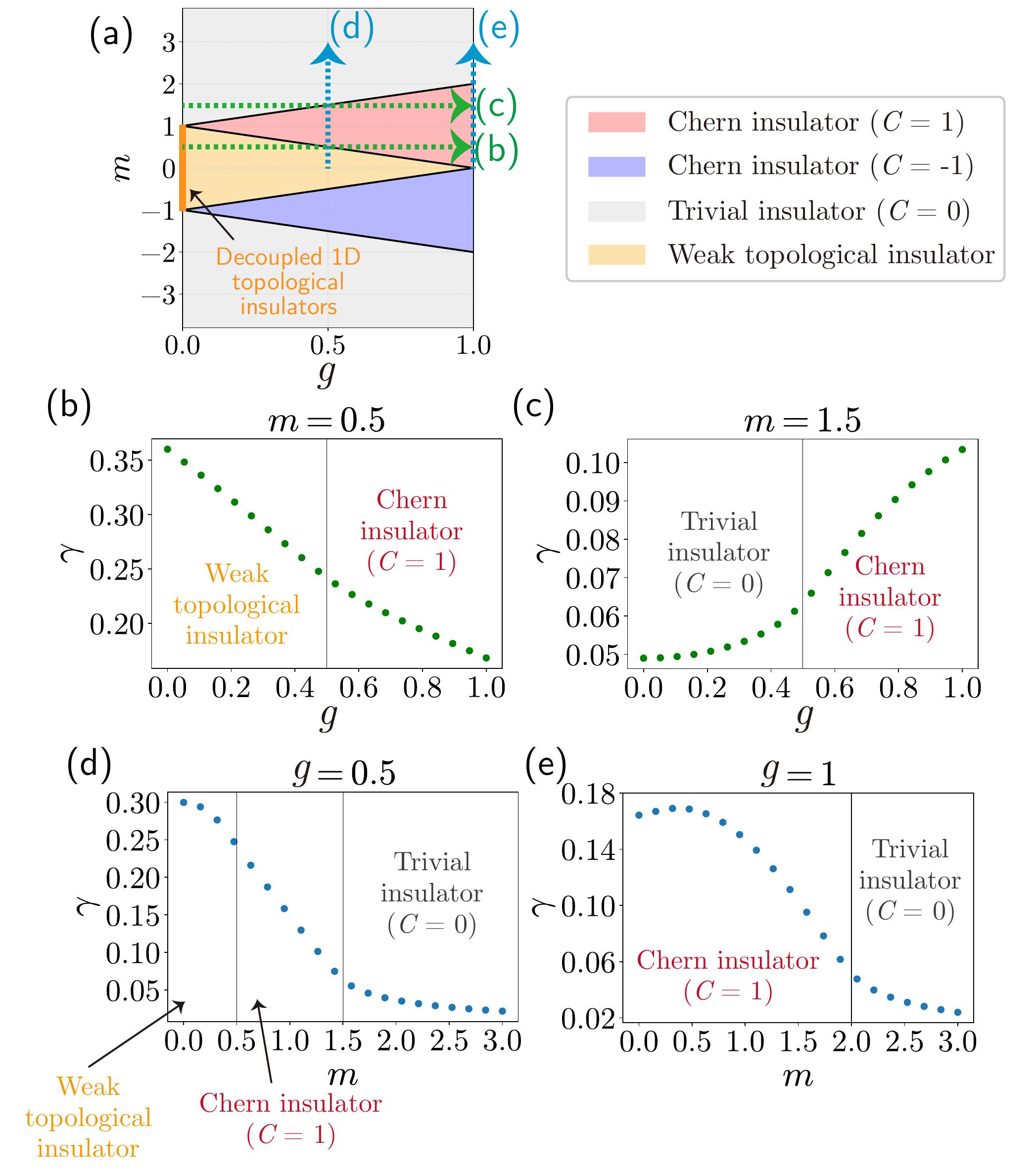}
	\caption{(a)~Phase diagram of the model in Eq.~\eqref{eq:SM_model} and the path along which the surface energy is calculated. (b) and (c)~Surface energy $\gamma$ for the (10) surface of the model versus $g$ for (b) $m=0.5$ and (c) $m=1.5$. (d) and (e)~Surface energy $\gamma$ for the (10) surface of the model versus $m$ for (d) $g=0.5$ and (e) $g=1$. The parameter is set to $v=0.7$, and the width of the strip is 40.}
	\label{fig:phase_diagram_surface_energy}
\end{figure} 

We illustrate how the boundary states of 1D topological insulators evolve into those of the Chern insulator. 
For $g=0$ and $|m|<1$, the model corresponds to decoupled 1D topological insulators hosting boundary states~[Fig.~\ref{fig:phase_diagram}(b)]. Although the Chern number vanishes in the region $|m|<-g+1$, boundary states originating from the limit of $g=0$ persist as in-gap states~[Fig.~\ref{fig:phase_diagram}(c)]. We denote this region as ``weak topological insulator'' in Fig.~\ref{fig:phase_diagram}(a).
Increasing $g$ while keeping $m$ constant causes the band gap to close at the parameters satisfying Eqs.~\eqref{eq:SM_gapclose2} or \eqref{eq:SM_gapclose3}~[Fig.~\ref{fig:phase_diagram}(d)]. 
A further increase in $g$ drives the system into the Chern insulator phase~[Fig.~\ref{fig:phase_diagram}(e)].   

\section{Surface energy in the topological phase transitions}\label{appendix: surface_energy_TPT}

In this section, we investigate the dependence of the surface energy $\gamma$ of the (10) surface on $g$ and $m$ in the tight-binding model defined by Eq.~\eqref{eq:SM_model}. 
We calculate $\gamma$ along the four paths shown in Fig.~\ref{fig:phase_diagram_surface_energy}(a), which traverse distinct topological phases.
By tracking the evolution of $\gamma$ along these paths, we analyze its behavior across the topological phase transitions.

Figure~\ref{fig:phase_diagram_surface_energy}(b) shows the variation in the surface energy $\gamma$ across the topological phase transition from the weak topological insulator to the Chern insulator. 
Here, $\gamma$ decreases as $g$ increases because the zero-energy boundary states ($E=0$) at $g=0$ evolve into the energetically favored boundary modes with linear dispersion, $E=vk_y$, as discussed in the main text. 
In contrast, Fig.~\ref{fig:phase_diagram_surface_energy}(c) shows the evolution of $\gamma$ across the topological phase transition from the trivial insulator to the Chern insulator. 
Here, $\gamma$ increases with $g$, contrary to the behavior observed in Fig.~\ref{fig:phase_diagram_surface_energy}(b).  This difference arises because the zero-energy boundary states do not appear at $g=0$ for $m=1.5$, unlike the case for $m=0.5$. Since the formation of chiral boundary states entails a higher energy cost than the absence of boundary states, $\gamma$ is higher in the Chern insulator than that in the trivial insulator.

Next, we investigate the dependence of the surface energy $\gamma$ on $m$.
Figure~\ref{fig:phase_diagram_surface_energy}(d) shows that $\gamma$ decreases as $m$ increases for $g=0.5$. 
In this process, the topological phase transition occurs from the weak topological insulator to the Chern insulator, followed by the transition to the trivial insulator.
This result indicates that while the chiral boundary states are energetically favored over the zero-energy boundary states, they incur a higher energy cost than the absence of boundary states.
Figure~\ref{fig:phase_diagram_surface_energy}(e) shows the dependence of $\gamma$ on $m$ for $g=1$. This result confirms that the trivial insulator is favored over the Chern insulator in terms of the surface energy, consistent with the behavior observed in Fig.~\ref{fig:phase_diagram_surface_energy}(c).

\section{Interfacial instability with negative stiffness}\label{appendix: interfacial_instability}
\begin{figure}
\includegraphics[width=1.\columnwidth]{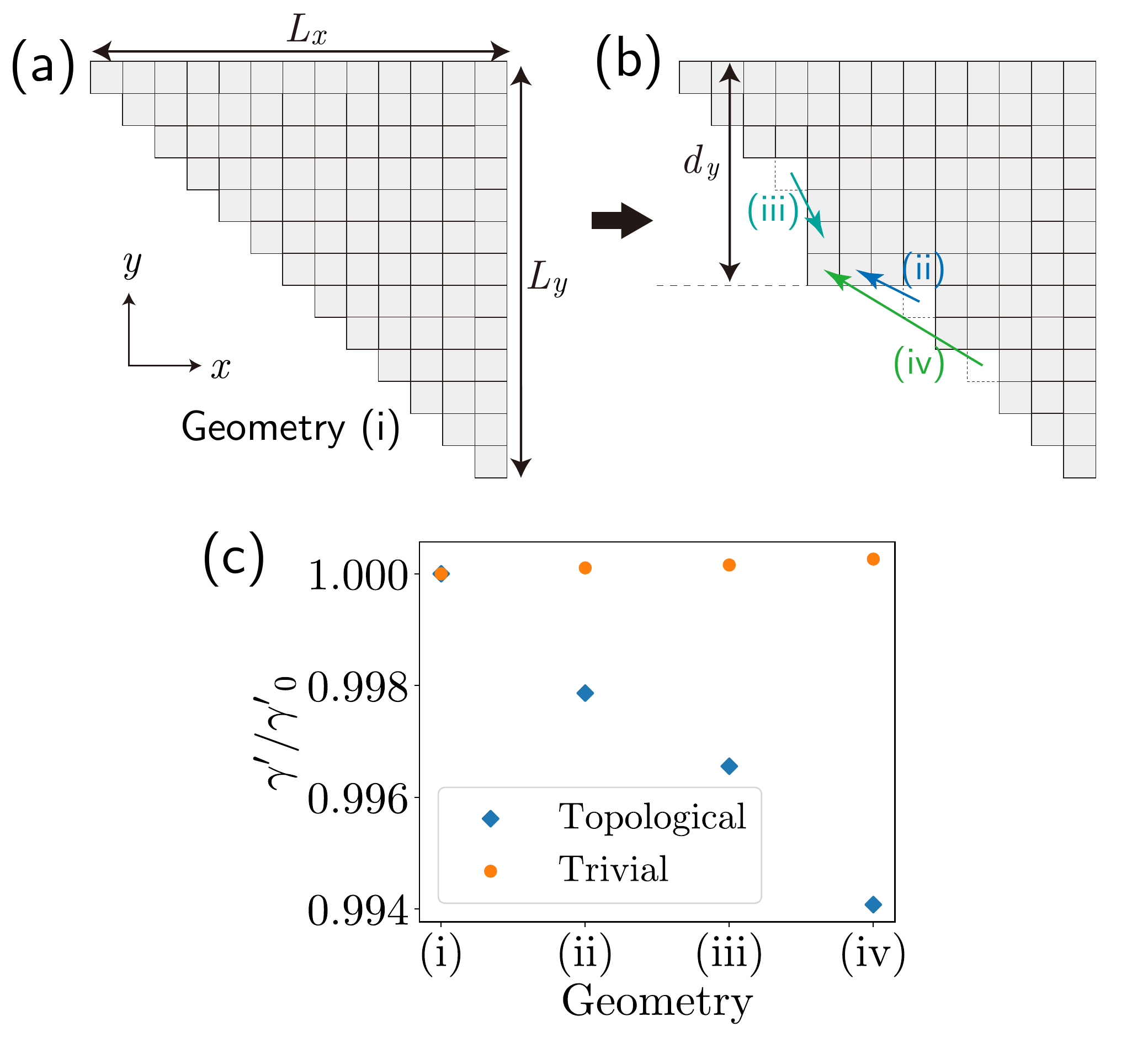}
\caption{(a)~The system under the full OBC with the ($10$), ($01$), and ($\bar{1}\bar{1}$) surfaces [geometry (i)]. (b)~ Morphological evolution from geometry (i). The atomic displacements shown in (ii), (iii), and (iv) occur on the ($\bar{1}\bar{1}$) surface.
(c)~The ratio of the surface energy ${\gamma}'$ of the Chern insulator in Eq.~\eqref{eq:Chern_model} to the reference surface energy ${\gamma}'_{0}$ under the full OBC with the geometries (i-iv).
Here, ${\gamma}'_{0}$ is the surface energy with the geometry (i). 
Geometries (ii), (iii), and (iv) are the structures obtained from the geometry (i) after the atomic displacements (ii), (iii), and (iv) in (b), respectively. 
The results for the topological and trivial phases are obtained for the parameters $m=0.8$ and $m=4$ in the model~[Eq.~\eqref{eq:Chern_model}], respectively~($v=0.7$, $L_x=30$, $L_y=30$, and $d_y=15$). 
}\label{fig:energy_change}
\end{figure}

To clarify the distinct behaviors between positive and negative stiffness, we investigate the change in the surface energy associated with a morphological evolution from the ($\bar{1}\bar{1}$) surface to a bumped surface~[Figs.~\ref{fig:energy_change}(a) and \ref{fig:energy_change}(b)]. 
We first consider a geometry under the full OBC with the (10), (01), and ($\bar{1}\bar{1}$) surfaces, which we refer to as geometry (i)~[Fig.~\ref{fig:energy_change}(a)]. 
We calculate the following surface energy for geometry (i):
\begin{align}\label{eq:average_surface_energy}
    {\gamma}' := \frac{E_{\rm OBC}-N_{\rm uc}E_{\rm bulk}}{L_x+L_y+\sqrt{L_x^2+L_y^2}},
\end{align}
where $L_x$ ($L_y$) is the system size along the $x$ ($y$) direction, $N_{\rm uc}$ is the number of unit cells in the geometry (i), and $E_{\rm bulk}$ is given by Eq.~\eqref{eq:bulk_enrgy}.
Here, $E_{\rm OBC}$ is the energy for the geometry (i) under the full OBC, which is defined by
\begin{align}
    E_{\rm OBC} := \sum_{n} E^{(n)}_{\rm OBC}\Theta \left( -E^{(n)}_{{\rm OBC}} \right),
\end{align}
where $E^{(n)}_{\rm OBC}$ is the $n$th energy eigenvalue under the full OBC. 

We consider the atomic displacements on the ($\bar{1}\bar{1}$) surface, labeled as (ii), (iii), and (iv) in Fig.~\ref{fig:energy_change}(b). The displacement (ii) occurs first, followed by (iii), and then by (iv). We refer to the geometries obtained after these displacements (ii), (iii), and (iv) as geometries (ii), (iii), and (iv), respectively. Figure~\ref{fig:energy_change}(c) shows the surface energy $\gamma'$ in Eq.~\eqref{eq:average_surface_energy} for the Chern insulator in Eq.~\eqref{eq:Chern_model} in the geometries (i)-(iv). 
In the topological phase~($m=0.8$), the surface energy $\gamma'$ decreases as the morphological evolution progresses, whereas in the trivial phase~($m=3$), $\gamma'$ increases during this process. 
The contrasting behaviors reflect the negative stiffness of the Chern insulator and positive stiffness of the trivial insulator. 

\section{Derivation of the interface growth equation}\label{appendix: derivation_interfaace_growth}
Here, we derive the equation describing the growing interface between a solid and its surrounding environment, such as vapor or liquid phases, following a well-established discussion of interface growth driven by a thermodynamic driving force~\cite{saito1996book, Pimpinelli_Villain_1998}. 
Since we consider 2D systems in this work, we focus on an interface $h(x,t)$ that depends on the position $x$ and time $t$.  
Taking the energy of the entire system in the environmental phase as the reference, the Gibbs free energy $\mathcal{G}$ is given by
\begin{align}\label{eq:free_energy}
    \mathcal{G}[h]= \int g(x,h, p, q)dx,  
\end{align}
where $p:=\partial_x h$, $q:=\partial_{xx}h$, and $g(x,h, p, q)$ is the free energy density, 
\begin{align}
    g=-n_s\Delta {\mu}  h + \gamma(p)  \sqrt{1+p^2}+\frac{\rho}{2}q^2. 
\end{align}
Here, $n_s$ is the particle density of the solid, $\Delta \mu$ ($>0$) is the difference in chemical potential between the environment and the solid phases, and $\gamma(p)$ is the surface energy.  
The third term, $\rho q^2/2$, represents the elastic free energy of an undulating solid surface, where $\rho$ is a coefficient related to the curvature rigidity, which reflects the energetic cost of forming a corner on the surface~\cite{landau2012theory, Stewart1992, Pimpinelli_Villain_1998, Nelson2004}.

When the interface growth is governed by interfacial kinetics, the interface velocity $V$ is proportional to the thermodynamic driving force, which is given by the variational derivative of the free energy, $-\delta \mathcal{G}/\delta h$:
\begin{align}
    V = {K}v_s \left(- \frac{\delta \mathcal{G}}{\delta h} \right),
\end{align}
where $K$ is the kinetic growth coefficient, and $v_s (=1/n_s)$ is the volume per particle.  
Since the variational derivative of the free energy is given by
\begin{align}
    \frac{\delta \mathcal{G}}{\delta h} = \frac{\partial g}{\partial h} -\frac{\partial}{\partial x} \frac{\partial g}{\partial p}+\frac{\partial^2}{\partial x^2}\frac{\partial g}{\partial q},
\end{align}
the interface velocity can be written as
\begin{align}\label{eq:growth_speed}
    V=K \biggl[\Delta \mu +v_s\frac{\partial}{\partial x}\frac{\partial }{\partial p}\left( \gamma(p)  \sqrt{1+p^2} \right)- v_s\rho \frac{\partial^4 h}{\partial x^4} \biggr]. 
\end{align}
Since the surface energy $\gamma(\theta)$ in Eq.~\eqref{eq:broken_bond_ec} is a function of the inclination angle $\theta$, we rewrite the second term of the right-hand slide of Eq.~\eqref{eq:growth_speed} by introducing the radius of curvature $R$ and the line element of the surface $s$:
\begin{align}
    \frac{\partial}{\partial x}\frac{\partial }{\partial p}\left( \gamma (p) \sqrt{1+p^2} \right)=&\frac{\partial s}{\partial x} \frac{\partial \theta}{\partial s} \frac{\partial }{\partial \theta}\left( \frac{\partial \theta}{\partial p } \frac{\partial}{\partial \theta} \left( \frac{\gamma}{\cos \theta} \right) \right) \nonumber \\
     = & -\frac{\tilde{\gamma}}{R},
\end{align}
where we use the relations $p=\tan \theta$, $\partial \theta/\partial s=-1/R$, and $\partial s /\partial x = 1/\cos \theta$, and $\tilde{\gamma}$ is the surface stiffness defined in Eq.~(\ref{eq:stiffness}). 
For a slightly tilted surface with $p\ll 1$, the curvature can be approximated as $1/R\simeq -q$, and thus Eq.~\eqref{eq:growth_speed} becomes
\begin{align}\label{eq:velocity_interface}
    V \simeq K \left( \Delta \mu +v_s \tilde{\gamma}q- v_s\rho \frac{\partial^4 h}{\partial x^4} \right).
\end{align}

We now incorporate a nonlinear effect into the interface growth equation, as such an effect naturally arises during interface evolution.
According to Ref.~\cite{Kardar1986}, the interface growth in the presence of the nonlinear contribution is given by 
\begin{align}
    \frac{\partial h}{\partial t}\simeq V\left( 1+\frac{p^2}{2}\right),
\end{align}
for a slightly tilted surface with $p\ll 1$. 
Substituting Eq.~\eqref{eq:velocity_interface} into this expression yields 
\begin{align}
     \frac{\partial h}{\partial t} \simeq K\left[\Delta \mu+ v_s \tilde{\gamma}q+\frac{\Delta \mu p^2}{2} - v_s\rho \frac{\partial^4 h}{\partial x^4} \right],
\end{align}
where we have neglected higher-order terms such as $p^2q$ and $p^2 \partial_{x}^4h$.
By applying a Galilean transformation $h \rightarrow h+K\Delta\mu t$ and introducing the coefficients $\nu:=Kv_s \tilde{\gamma}$, $\kappa:=K\Delta\mu/2$, and $\eta:=Kv_s\rho$, the equation takes the form of Eq.~\eqref{eq:interface_growth} in the main text. 


%

\end{document}